\documentclass[aps,prl,twocolumn,superscriptaddress,longbibliography]{revtex4-2}
\usepackage[normalem]{ulem}

\usepackage{amsmath,amssymb,amsfonts}
\usepackage{graphicx}
\usepackage{bm}
\usepackage{booktabs}
\usepackage{hyperref}
\usepackage[capitalize]{cleveref}
\usepackage{xcolor}

\newcommand{\X}{\mathcal{X}}
\newcommand{\Y}{\mathcal{Y}}
\newcommand{\KL}[2]{D(#1\,\|\,#2)}
\newcommand{\MMC}{\mathrm{MMC}}
\newcommand{\Lc}{\mathcal{L}_G(\X)}
\DeclareMathOperator*{\argmax}{arg\,max}

\begin{document}

\title{Minimax entropy production for arbitrary processes}

\author{Mahran Yousef}
\affiliation{Independent}

\author{Ben Ansbacher}
\affiliation{Santa Fe Institute, Santa Fe, NM 87501, USA}

\author{David H.\ Wolpert}
\affiliation{Santa Fe Institute, Santa Fe, NM 87501, USA}
\affiliation{Complexity Science Hub, Vienna, Austria}
\affiliation{Arizona State University, Tempe, AZ 85287, USA}
\affiliation{International Centre for Theoretical Physics, Trieste, Italy}

\date{\today}
\begin{abstract}

For any physical process, varying the initial distribution over its states will change the resulting expected entropy production (EP). An initial distribution that minimizes the EP of a particular process is called the \emph{prior} of that process. Varying the microphysical details of the process will change its prior. We consider the following question: if an engineer wants to design a system to implement a given map, what prior should they design into the system to minimize the EP for the worst-case actual distribution? We refer to the resulting optimal worst-case value as the \emph{minimax EP}, and investigate how it depends \emph{only} on the dynamics of the process. We derive the closed-form solution to this minimax problem when the map is deterministic. We also show that for stochastic maps, a worst-case actual distribution can always be chosen to be a point mass on a single state, and that the associated minimax prior resembles a Boltzmann-Gibbs equation and can be found by a globally convergent numerical algorithm. We further show that the minimax EP is the maximum uncertainty about the input that remains after observing the output. Finally, we extend the analysis to the continuous-time regime, showing that the minimax instantaneous EP diverges, while the corresponding finite-time minimax EP has a universal short-time scaling set by the largest escape rate of the dynamics.
\end{abstract}

\maketitle

\textit{Introduction--- }
Consider a physical system with a finite state space $\X$,
and a physical process that acts over a fixed non-infinitesimal time interval
$[0,t]$ on that system. Any such process induces a conditional distribution
$G$ over $\X$, sending an initial distribution $p$ to a final distribution
$Gp$. For now, we leave the time interval implicit and write the resulting
finite-time map simply as $G$.
Several important thermodynamic properties of such systems have the form
\begin{equation}
C(G,f,p)=S(Gp)-S(p)+\sum_x f(x)p(x),
\label{eq:general-cost}
\end{equation}
where $S(p)=-\sum_xp(x)\log p(x)$ is the Shannon entropy. In general, the
function $f$ can reflect both a thermodynamic property of interest and the
microscopic physics of the system. An important example is entropy production
(EP), where $f(x)$ is the expected entropy flow out of the system conditioned
on starting in state $x$~\cite{KolchinskyWolpert2017,KolchinskyWolpert2021}.

In many applications, however, the input distribution $p$ is not known when
the physical system is designed. An engineer may know the map $G$ that a
device must implement, but not the distribution of inputs it will actually
encounter. Computers are an example of this, where a variety of input distributions are induced by different users. The problem we are interested in is therefore: how should the
physical implementation of $G$ be chosen so that the extra thermodynamic cost
caused by this uncertainty is as small as possible in the worst case?

Engineers routinely use minimax or related worst-case design for this kind of
problem. When the disturbances, noise, or inputs are not known in advance,
one chooses the design that minimizes the largest possible cost over the
allowed possibilities. This basic idea appears throughout optimization,
control, and signal-processing design
~\cite{BenTalNemirovski1998,DoyleEtAl1989,Cheng1999}. We are motivated by the
same idea here. We ask which physical implementation of a fixed map $G$
minimizes the largest extra thermodynamic cost caused by running the system
on an input distribution different from the one it was designed for. 

\emph{Mismatch cost--- } A growing body of work has characterized how a cost function like \eqref{eq:general-cost} depends on
the initial distribution $p$ using the Mismatch Cost (MMC) formalism ~\cite{KolchinskyWolpert2017,WolpertReview2019,KolchinskyWolpert2020Turing,WolpertKolchinsky2020Circuits,KolchinskyWolpert2021,RiechersGu2021,Wolpert2020Biology,WolpertCircuits2020,OuldridgeWolpert2023,ManzanoEtAl2024,WolpertEtAl2024,YadavWolpert2025,YadavYousefWolpert2025,YadavCaravelliWolpert2026,SchaeferAnsbacher2026}. To do this, we consider the distribution for which \eqref{eq:general-cost} is a minimum. This distribution $q:= \text{argmin}_r C(G,f,r)$ is called the \emph{prior}. The cost then can be decomposed as 
\begin{equation}
    C(G,f,p) = \underbrace{D(p\|q)-D(Gp\|Gq)}_{\MMC(p)} + R,
\end{equation}
where $\MMC(p)$ is the Mismatch Cost and $R=C(G,f,q)$ is the residual cost: the cost incurred by using the prior $q$ (the full general definition of mismatch cost requires the discussion of islands, which we include below). From the data-processing inequality for KL divergence, the MMC is non-negative, and therefore sets a lower bound on any cost function of the form in \eqref{eq:general-cost}, including EP. It has been shown that MMC can take up a significant fraction of the total EP in various physical systems \cite{SchaeferAnsbacher2026, YadavCaravelliWolpert2026}. Crucially, the prior $q$ depends on the microphysics---like the entropy-flows described by $f$---of the process and therefore its physical implementation. Changing $q$ thus changes the physical implementation of the map $G$. 

To discuss MMC in full generality, it is important to define the \emph{islands} of
$G$. Intuitively, islands are `walled off' subsystems of the dynamics that become relevant in modular systems like computers and circuits \cite{OuldridgeWolpert2023, YadavYousefWolpert2025}, and become necessary to capture the cost's full dependence on initial distribution. Formally, they account for the fact that the prior that minimizes \eqref{eq:general-cost} is not always unique. We first define the following relation on the input states: two inputs
$x$ and $x'$ are related if there is at least one output $y$ that both inputs
can produce with nonzero probability,
\[
x \sim x'
\quad\Longleftrightarrow\quad
\exists y:\; G(y\mid x)>0
\;\text{and}\;
G(y\mid x')>0 .
\]
The islands of $G$ are the equivalence classes generated by the transitive
closure of this relation. In other words, two inputs belong to the same
island if they can be connected by a chain of input states in which each
consecutive pair shares at least one possible output~\cite{WolpertReview2019}.

We denote an island by $c\in\Lc$ and the set of outputs that can be produced
from inputs in island $c$ by $\Y_c$. The islands partition the input space so
that inputs belonging to different islands never produce the same
output. In the case of a deterministic map---a map that always produces one
output for a given input---the islands are the preimages of that map.

For an input distribution $p$, let $p(c)=\sum_{x\in c}p(x)$ be the total
probability assigned to island $c$, and let
$p^c(x)=p(x)/p(c)$ for $x\in c$ and $p(c)>0$ be the conditional probability
of state $x$ in island $c$. Let $G_c$ denote the same map restricted to
inputs in $c$. For fixed $G$ and $f$, each island has a conditional
distribution $q^c$ that minimizes the restricted cost
\begin{equation}
C_c(G,f,r)=S(G_cr)-S(r)+\sum_{x\in c}f(x)r(x),
\end{equation}
over distributions $r$ on $c$. We call $q^c$ the \emph{prior within island
$c$}. The total cost then decomposes as
\begin{equation}
C(G,f,p)=\sum_{c\in\Lc}p(c)
\left[R_c+\MMC_c(p^c,q^c,G_c)\right],
\label{eq:cost-decomp}
\end{equation}
where $R_c=C_c(G,f,q^c)$ is the residual cost in island $c$, which is the
minimum cost produced by that process, and
\begin{equation}
\MMC_c(r,q,G_c)=\KL{r}{q}-\KL{G_cr}{G_cq}\ge0
\label{eq:mc}
\end{equation}
is the MMC within that island
~\cite{KolchinskyWolpert2021,YadavCaravelliWolpert2026}. We call the set of priors over the islands
$\bm q=\{q^c\}_{c\in\Lc}$ and define the total mismatch cost
\begin{equation}
\MMC(p,\bm q,G):=
\sum_{c\in\Lc}p(c)\MMC_c(p^c,q^c,G_c).
\label{eq:total-mmc}
\end{equation}
The full prior is not unique, and is given by $q=\sum_c \eta_c q^c $ where the $\eta_c>0$ are arbitrary and subject to $\sum_c \eta_c=1$. The total mismatch cost is independent of these island weights given by $\eta_c$.


Different physical implementations of the same $G$ can have different priors within their islands. The engineer chooses the implementation, and therefore its priors, before knowing the actual input distribution $p$. We focus on the mismatch term because, after choosing a desired full-support prior $q$, one can in principle construct a quasistatic implementation of the same map $G$ for which $q$ is mapped to $Gq$ with vanishing residual EP. The remaining additional cost when the actual input distribution differs from the chosen prior is then the MMC.

We can now state the minimax problem precisely. For a fixed map $G$, we ask
which choice of island priors makes the largest possible mismatch cost over
all input distributions as small as possible. The actual input distribution
may be any $p\in\Delta(\X)$, where $\Delta(\X)$ denotes the set of
probability distributions over $\X$. We therefore define
\begin{equation}
F_G(\bm q):=\max_{p\in\Delta(\X)}\MMC(p,\bm q,G),
\qquad
F_G^*:=\min_{\bm q}F_G(\bm q),
\label{eq:minimax}
\end{equation}
where each $q^c\in\Delta(c)$. We first solve this problem for deterministic maps. We then generalize the
result to arbitrary stochastic maps, characterize the minimizing priors, and
give an algorithm for finding them. Finally, we ask the corresponding question in continuous time. 

\textit{Deterministic maps--- }
Let $G$ be deterministic, so that every input $x$ is mapped to a single
output $g(x)$. Thus
\[
G(y|x)=\mathbf 1\{y=g(x)\},
\]
where $\mathbf 1\{y=g(x)\}$ equals $1$ when $y=g(x)$ and $0$ otherwise. Two inputs can
produce a common output exactly when they are mapped to the same $y$. The islands are therefore the preimages of $G$
\begin{equation}
I_y=\{x:g(x)=y\}.
\end{equation}
For a candidate conditional prior \(q^y\) on island \(I_y\), the worst-case input distribution is concentrated on the state assigned the smallest probability by \(q^y\). Therefore,
\begin{equation}
F_y(q^y)=\max_{x\in I_y}[-\log q^y(x)].
\end{equation}
The smallest possible worst-case mismatch cost within \(I_y\) is obtained by choosing \(q^y\) to be the uniform distribution over that island, which gives
\begin{equation}
F_y^*=\log |I_y|.
\end{equation}
The global worst-case input can be concentrated on any island, so the minimax value for the full deterministic map is the largest of these islandwise values:
\begin{equation}
F_G^*=\max_y F_y^*
=\log\max_y|I_y|.
\label{eq:det-result}
\end{equation}
Let $L_{\max}=\max_y|I_y|$. For every island with \(|I_y|=L_{\max}\), the prior must be uniform in order to attain the global minimax value. For a smaller island, uniformity within that island is not necessary: any prior satisfying
\begin{equation}
\min_{x\in I_y}q^y(x)\ge \frac{1}{L_{\max}}
\end{equation}
has a worst-case mismatch cost no larger than \(F_G^*\).  This result matches the minimax result found for local, periodic deterministic finite automata by
Ouldridge and Wolpert~\cite{OuldridgeWolpert2023}.

A clear application of this result is to computers, since oftentimes they encounter a variety of different input distributions. The stochastic thermodynamics of computation has been studied at length using the MMC \cite{KolchinskyWolpert2020Turing, YadavCaravelliWolpert2026, SchaeferAnsbacher2026, WolpertKolchinsky2020Circuits}. In particular, we can consider how the result for deterministic maps depends on the physical description of the computation. The same map, a boolean function for example, can be treated as one coarse-grained operation or as a sequence of individual boolean gate operations. These two descriptions need not have the same minimax mismatch cost. 

To illustrate this,  consider a Boolean circuit of AND gates arranged as binary tree of depth $d$. It has $2^d$ input bits and $2^d-1$ gates. If the whole
circuit is treated as one deterministic map from all input bits to the final
output bit, the output-$0$ island contains every input except the all-ones
input. Hence
\begin{equation}
F^*_{\rm map}(d)=\log(2^{2^d}-1)\approx2^d\log2.
\label{eq:tree-map}
\end{equation}
If instead each AND gate is treated as a separate physical operation, its output-$0$ island is
$\{00,01,10\}$, so each gate's minimax mismatch cost is $\log3$, giving the circuit a lower bound on its minimax MMC  of 
\begin{equation}
F^*_{\rm gates}(d)=(2^d-1)\log3.
\label{eq:tree-gates}
\end{equation}
A more detailed discussion of this example, together with an illustration of the binary AND tree, is given in Supplemental Material Sec.~\ref{S:sec:circuit}~\cite{SupplementalMaterial}.

\begin{figure}[t]
\centering
\includegraphics[width=0.84\columnwidth]{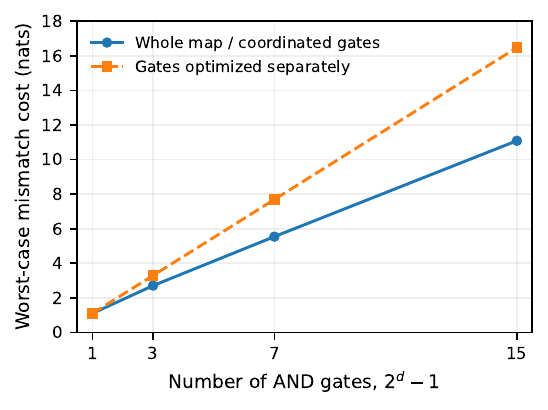}
\caption{Minimax mismatch cost of a depth-$d$ AND tree ($d=1,\dots,4$),
computed as one Boolean map using Eq.~\eqref{eq:tree-map} and gate by gate using
Eq.~\eqref{eq:tree-gates}. Both grow linearly with the number of gates, with
different coefficients because they describe different physical
implementations. The gate-by-gate curve is the sum of local gate mismatch
costs and is a lower bound when correlations with the rest of the circuit are
included~\cite{YadavYousefWolpert2025}.}
\label{fig:andtree}
\end{figure}

The two minimax mismatch costs are compared in Fig.~\ref{fig:andtree}.
The figure shows that both costs grow approximately linearly with the number
of gates, but with different coefficients. Even for the same logical computation, the minimax mismatch
cost depends on how the computation is physically decomposed. For the
gate-resolved description, Eq.~\eqref{eq:tree-gates} gives the sum of the
local gate mismatch costs and therefore a lower bound on the total circuit
mismatch cost (correlations between a gate and the rest of the circuit can
add mutual-information terms~\cite{YadavYousefWolpert2025}).

\textit{Stochastic maps--- }
We now turn to stochastic maps, for which one input can produce several
possible outputs. As in the deterministic case, we consider one island at a
time. Let $c$ be an island, let $q^c$ be a candidate prior on that island,
and let $\delta_x$ denote the point-mass distribution on $x$. For an input
concentrated entirely on state $x$, Eq.~\eqref{eq:mc} gives
\begin{equation}
\MMC_c(\delta_x,q^c,G_c)
=-\log q^c(x)-\KL{G(\cdot|x)}{G_cq^c}.
\end{equation}
The first term is large when the prior assigns a small probability to state $x$, while the KL term is large when the output distribution produced by \(x\) differs strongly from the overall output distribution produced by the prior. For
convenience, define
$d_c(x,q^c):=\KL{G(\cdot|x)}{G_cq^c}$.

Guided by the deterministic case, we ask whether an input distribution spread
over several states can ever produce a larger mismatch cost than any point mass input distribution on a single state. We show in Supplemental Material
Sec.~\ref{S:sec:pointmass} that it cannot. For a fixed prior $q^c$, let
$F_c(q^c)$ denote the largest mismatch cost that can occur within island $c$.
A worst-case input is therefore always a point mass:
\begin{align}
F_c(q^c)
&:=\max_{r\in\Delta(c)}\MMC_c(r,q^c,G_c)
\nonumber\\
&=\max_{x\in c}
\left[-\log q^c(x)-d_c(x,q^c)\right].
\label{eq:pm}
\end{align}

The next question is which prior makes this worst-case cost as small as
possible. We show in Supplemental Material Sec.~\ref{S:sec:optimal} that the minimizing prior is unique, has full support and is given by~\cite{SupplementalMaterial,NedicEtAl2025}:
\begin{align}
q_c^*
&=\argmax_{r\in\Delta(c)}
H_r(X|Y,J=c),
\nonumber\\
F_c^*
&:=\min_{q^c\in\Delta(c)}F_c(q^c)
=\max_{r\in\Delta(c)}
H_r(X|Y,J=c).
\label{eq:local-dual}
\end{align}
Intuitively, the minimax cost is equal to the largest uncertainty about the input that can remain after observing the output. The larger this uncertainty, the larger the worst-case mismatch cost.

The same result gives an explicit condition for the minimax prior. At
$q_c^*$, the mismatch costs of all point-mass inputs in the island are equal,
as shown in Supplemental Material Sec.~\ref{S:sec:optimal}. This gives
\begin{equation}
q_c^*(x)=
\frac{\exp[-d_c(x,q_c^*)]}
{\sum_{z\in c}\exp[-d_c(z,q_c^*)]},
\qquad x\in c.
\label{eq:fp}
\end{equation}
This is an implicit equation for \(q_c^*\) and resembles a Boltzmann-Gibbs distribution. Unlike an ordinary Boltzmann-Gibbs distribution, however, the quantity in the exponential also depends on the prior \(q_c^*\) itself.

Combining the island-wise results gives
\begin{equation}
F_G(\bm q)=\max_{c\in\Lc}F_c(q^c),
\qquad
F_G^*=\max_{c\in\Lc}F_c^*.
\label{eq:global-F}
\end{equation}
Choosing $q_c^*$ in every island therefore gives a globally minimizing
collection of priors. However, this is not the only possible minimizing
collection. Only islands whose local minimax value equals the global value
$F_G^*$ must use their minimizer priors $q_c^*$; priors in the other islands may be different from their respective minimizer priors as long as their worst-case cost does not exceed $F_G^*$. The
complete set of global minimizers is characterized in Supplemental Material
Sec.~\ref{S:sec:optimal}.

Since outputs produced by different islands do not overlap, observing the output also identifies the island. The remaining uncertainty about the input is therefore the weighted average of the uncertainty within each island. Maximizing over the input distribution gives
\begin{equation}
F_G^*=\max_{p\in\Delta(\X)}H_p(X|Y).
\label{eq:dual}
\end{equation}
A proof likewise can be found in Supplemental Material
Sec.~\ref{S:sec:optimal}. The maximization of $H(X|Y)$ has previously appeared in information theory
as the maximum-equivocation or commitment-capacity problem
~\cite{WinterNascimentoImai2003,NedicEtAl2025}; here we show that the same
quantity is exactly the minimax mismatch cost of an arbitrary physical process.
For deterministic maps, Eq.~\eqref{eq:dual} reduces to
Eq.~\eqref{eq:det-result}.

Additionally, Eq.~\eqref{eq:fp} provides a direct numerical method for finding
$q_c^*$. Starting from any full-support $q_k^c$, define
\begin{equation}
q_{k+1}^c(x)=
\frac{\exp[-d_c(x,q_k^c)]}
{\sum_{z\in c}\exp[-d_c(z,q_k^c)]}.
\label{eq:algorithm}
\end{equation}
This algorithm converges to \(q_c^*\) from any full-support initial distribution under the conditions established in Refs.~\cite{Hayashi2025,NedicEtAl2025}. Supplemental Material Sec.~\ref{S:sec:algorithm} gives the derivation, convergence conditions, and computable lower and upper bounds on the minimax value. 

Lastly, it is important to point out that the minimax prior should not be confused with a stationary distribution. One might think intuitively that the stationary distribution of G, $\pi_{ss}$, which satisfies
\begin{equation}
G\pi_{ss}=\pi_{ss},
\end{equation}
might be a good candidate for a minimax prior. However, this is not true in general. In the Supplemental Material
Sec.~\ref{S:sec:ring} we give an example of a driven three-state cycle that illustrates how the costs incurred by the minimax prior and the stationary distribution differ.

\textit{Continuous time--- }
We now ask the same minimax question for a continuous-time process.
We begin without assuming Markovian dynamics. Let $P_t$ be a
differentiable family of stochastic maps with $P_0=I$, and define
\begin{equation}
A:=\left.\frac{dP_t}{dt}\right|_{t=0}.
\end{equation}
Thus, for an initial distribution $p$, the initial change in the distribution
is $\dot p_0=Ap$. We do not assume that $P_t=e^{tA}$.

Let $f_t(x)$ be the expected entropy flow out of the system during the
time interval $[0,t]$ when the system starts in state $x$, with $f_0=0$.
The total EP (denoted $W_d$ for its connection to dissipated work) accumulated over this interval is
\begin{equation}
W_d(t;p)=S(P_tp)-S(p)+p\cdot f_t.
\end{equation}
The EP rate at the initial time $t=0$ is
\begin{equation}
\sigma(p):=\left.\frac{d}{dt}W_d(t;p)\right|_{t=0}
=-(Ap)\cdot\ln p+p\cdot\dot f_0.
\label{eq:sigma}
\end{equation}
Let $q^\sharp$ be the initial distribution that minimizes the initial EP rate
$\sigma(p)$, and suppose it has full support. The condition for $q^\sharp$ to
minimize $\sigma(p)$ is
\begin{equation}
-\sum_yA_{yx}\ln q^\sharp(y)
-\frac{(Aq^\sharp)(x)}{q^\sharp(x)}
+\dot f_0(x)=\sigma(q^\sharp),
\label{eq:rate-prior}
\end{equation}
Using this condition, for any other full-support initial distribution $p$ we
obtain
\begin{equation}
\sigma(p)-\sigma(q^\sharp)
=-\left.\frac{d}{dt}\KL{P_tp}{P_tq^\sharp}\right|_{t=0}.
\label{eq:rate-mmc}
\end{equation}
This follows from Eqs.~\eqref{S:eq:sigma} and
\eqref{S:eq:rate-prior}, which give Eq.~\eqref{S:eq:rate-mmc} in the
Supplemental Material~\cite{SupplementalMaterial}.

This motivates defining the instantaneous mismatch-cost rate for any
full-support distributions $p$ and $q$ as
\begin{equation}
\dot{\MMC}_A(p,q)
:=-\left.\frac{d}{dt}\KL{P_tp}{P_tq}\right|_{t=0}\ge0.
\label{eq:general-rate-mmc}
\end{equation}
The rate depends only on the initial derivative $A=\dot P_0$, not on whether
the full dynamics form a Markov semigroup.

We can now consider the continuous-time version of the minimax question:
which choice of $q$ makes the largest possible instantaneous mismatch-cost
rate as small as possible? We define
\begin{equation}
\dot F^*(A):=
\inf_{q\in\Delta^\circ (\X)}
\sup_{p\in\Delta^\circ (\X)}\dot{\MMC}_A(p,q),
\end{equation}
where $\Delta^\circ (\X)$ is the set of full-support distributions. If $A$ is nontrivial, meaning that it has at least one positive
off-diagonal entry,
\begin{equation}
\dot F^*(A)=+\infty.
\label{eq:rate-infinite}
\end{equation}
To see why, consider a state $a$ with a positive escape rate and let the
initial distribution $p$ place almost all of its probability on $a$, while
assigning a small probability $\epsilon$ to all other states. Probability
then begins to flow out of $a$ at a finite rate, including into states whose
initial probabilities are only of order $\epsilon$. The instantaneous
mismatch rate contains logarithms of these small probabilities, producing a
term proportional to $\log(1/\epsilon)$. Therefore,
\begin{equation}
\dot{\MMC}_A(p,q)
\sim \lambda_a\log(1/\epsilon)
\end{equation}
as $\epsilon\to0^+$, where $\lambda_a=-A_{aa}$ is the initial escape rate from $a$.
This rate can be made arbitrarily large by making $p$ arbitrarily close to a
point mass. For every fixed full-support $p$ and $q$, however, the rate
remains finite. The proof is given in Supplemental Material
Sec.~\ref{S:sec:shorttime}.

A simple three-state nonequilibrium steady-state example is worked out in
Supplemental Material Sec.~\ref{S:sec:shorttime}. Supplemental
Fig.~\ref{S:fig:rate-divergence} shows that its exact instantaneous mismatch
rate follows the predicted logarithmic growth and increases without bound as
the input distribution approaches a point mass. This gives a direct numerical
example of the divergence in Eq.~\eqref{eq:rate-infinite}.

To better understand this singular instantaneous limit, we now ask how the
finite-time minimax mismatch cost behaves as the duration of the process
approaches zero. This also lets us distinguish the divergence of the
instantaneous mismatch-cost rate from the behavior of the accumulated mismatch
cost over a short but nonzero interval. To do this, we now specialize to a
continuous-time Markov chain with
\begin{equation}
P_t=e^{tK},
\end{equation}
where $K$ is a finite-state transition-rate matrix. Assume that $K$ is
irreducible, so every state can eventually reach every other state. For every
$t>0$, the finite-time map $P_t$ has one island, so our earlier result
Eq.~\eqref{eq:dual} gives
\begin{equation}
F_t^*=\max_p H_p(X_0|X_t).
\end{equation}
This means that over finite time intervals, the minimax mismatch cost is the largest
possible uncertainty about the initial state $X_0$ after observing the state
$X_t$. Let
\begin{equation}
\lambda_x:=-K_{xx},
\qquad
\lambda_{\max}:=\max_x\lambda_x,
\end{equation}
where $\lambda_x$ is the total rate at which probability leaves state $x$.
For a very short time $t$, a system starting in state $x$ remains there
with probability $1-\lambda_x t+O(t^2)$ and leaves with probability
$\lambda_x t+O(t^2)$. The uncertainty about the initial state therefore
comes mainly from these transitions. Since the entropy associated with an
event of probability of order $t$ scales as $t\log(1/t)$, for a fixed
full-support initial distribution $p$ we have
\begin{equation}
H_p(X_0|X_t)
=
\bar\lambda(p)t\log(\tau_0/t)+O(t),
\end{equation}
where $\tau_0>0$ is an arbitrary fixed reference time and
\begin{equation}
\bar\lambda(p):=\sum_xp(x)\lambda_x
\end{equation}
is the average escape rate under $p$. To maximize this short-time
uncertainty, $p$ should be arbitrarily close to a point mass on the state with
the largest escape rate. Therefore,
\begin{equation}
\lim_{t\to0^+}
\frac{F_t^*}{t\log(\tau_0/t)}
=\lambda_{\max}.
\label{eq:small-time-minimax}
\end{equation}
This further gives for small $t$ that $F_t^* \sim \lambda_{\max} t\log(\tau_0/t)$. Since $t\log(\tau_0/t)\to0$, the finite-time minimax mismatch cost itself
vanishes,
\begin{equation}
F_t^*\to0
\qquad (t\to0^+),
\end{equation}
while the cost per unit time satisfies
$F_t^*/t\sim\lambda_{\max}\log(\tau_0/t)$ and therefore diverges. The short-time minimax mismatch is dictated by the largest escape rate of the rate matrix. In particular, it decays to zero as $t\log(\tau_0/t)$, scaled by this largest escape rate.
Details can be found in Supplemental Material \ref{S:sec:shorttime}.

The minimax prior shows similar short-time behavior. Let
$S_{\max}:=\{x:\lambda_x=\lambda_{\max}\}$
be the set of states with the largest escape rate. The short-time result
implies that the average escape rate under the minimax prior approaches
$\lambda_{\max}$:
\begin{equation}
\sum_x q_t^*(x)\lambda_x\longrightarrow\lambda_{\max}.
\end{equation}
This can only happen if the total probability assigned to states with
smaller escape rates approaches zero. Therefore,
\begin{equation}
q_t^*(S_{\max})
=
\sum_{x\in S_{\max}}q_t^*(x)
\longrightarrow1
\qquad(t\to0^+).
\label{eq:prior-limit}
\end{equation}
Thus, for very short times, the minimax prior probabilities are concentrated
on states with the largest escape rate. If only one state has escape rate
$\lambda_{\max}$, then $q_t^*$ approaches a point-mass distribution on that
state. Details can be found in Supplemental Material \ref{S:sec:shorttime}.

\textit{Discussion.}---
For a finite stochastic map, the central result is
$F_G^*=\max_pH_p(X|Y)$. The minimax mismatch cost is therefore determined by
how much information about the input can remain hidden after the output is
observed. Within each island of the map there is one unique prior $q_c^*$ that minimizes that
island's worst-case cost, and the global value is set by the island with the
largest such minimum. The island priors resemble Boltzmann-Gibbs distributions and can be obtained with a convergent numerical algorithm. 

In continuous time, the minimax results differ. The worst instantaneous MMC rate is unbounded when the input approaches the boundary of the probability simplex, while the finite-time minimax cost is finite and its short-time behavior is controlled by the largest escape rate of the rate matrix. 

The minimax problem presented and solved in this Letter is of use in many physical systems where the initial distribution is unknown. Computers, circuits, and communication channels must all account for worst-case input distributions induced by different users or use-cases. The minimax priors and lower bounds on entropy production are informative for designers of these systems since one can theoretically choose a physical implementation that gives rise to the minimax prior, thereby minimizing the worst-case mismatch cost. 

Future studies can apply the minimax criterion to previous work on periodic mismatch cost \cite{OuldridgeWolpert2023, SchaeferAnsbacher2026, YadavCaravelliWolpert2026}, which applies to instances where the same map $G$ and same physical process is repeated over and over. Additionally, for a fixed $G$, one can classify the set of functions $f$ in \eqref{eq:general-cost} (if they exist) that give rise to the minimax prior, and discuss how that informs the physical implementation of $G$. 

\begin{acknowledgments}
\textit{D.H.W. was supported by the Santa Fe Institute and NSF award 2626044. B.A. and D.H.W. were supported by a research grant (FOCUSED, VIL73405) from Villum Fonden}
\end{acknowledgments}

\bibliographystyle{apsrev4-2}
\bibliography{bibliography}

\clearpage
\onecolumngrid
\setcounter{section}{0}
\setcounter{secnumdepth}{3}
\setcounter{equation}{0}
\setcounter{table}{0}
\renewcommand{\thesection}{S\Roman{section}}
\renewcommand{\theequation}{S\arabic{equation}}
\renewcommand{\thetable}{S\Roman{table}}
\renewcommand{\theHsection}{supp.\arabic{section}}
\renewcommand{\theHequation}{supp.\arabic{equation}}
\renewcommand{\theHtable}{supp.\arabic{table}}

\begin{center}
{\large\bfseries Supplemental Material for ``Minimax entropy production for arbitrary processes''}\par
\vspace{0.75em}
Mahran Yousef, Ben Ansbacher, \& David H.\ Wolpert
\end{center}
\vspace{0.75em}

This Supplemental Material gives the derivations used in the Letter. We use
$G$ for a finite stochastic map, $p$ for the actual input distribution, and
$\bm q=\{q^c\}$ for a collection of candidate island priors. For an island
$c$, $\Delta(c)$ denotes the set of probability distributions on $c$, and
$\Delta^\circ (c)$ denotes the distributions that assign positive probability
to every state in $c$.

\section{Stochastic islands and the cost decomposition}
\label{S:sec:islands}

Two inputs are in the same stochastic island when they can be connected by a
chain of input pairs that share at least one possible output. For an island
$c$, define the set of outputs that can be produced from that island by
\begin{equation}
\Y_c=\bigcup_{x\in c}\{y:G(y|x)>0\}.
\label{S:eq:footprint}
\end{equation}
By construction of the islands, the sets $\Y_c$ for different islands
are disjoint.

For an input distribution $p$, write
\begin{equation}
p(c)=\sum_{x\in c}p(x),
\qquad
p^c(x)=\frac{p(x)}{p(c)}
\quad\text{when }p(c)>0.
\end{equation}
Let $J$ denote the island label. Both the input $X$ and the output $Y$
determine $J$, so
\begin{align}
S(p)&=H(J)+\sum_cp(c)S(p^c),\nonumber\\
S(Gp)&=H(J)+\sum_cp(c)S(G_cp^c).
\label{S:eq:entropy-islands}
\end{align}
Subtracting the two equations cancels $H(J)$. The linear term in
Eq.~\eqref{eq:general-cost} also separates over islands. Hence
\begin{equation}
C(G,f,p)=\sum_cp(c)C_c(G,f,p^c),
\label{S:eq:cost-island-sum}
\end{equation}
where
\begin{equation}
C_c(G,f,r)=S(G_cr)-S(r)+\sum_{x\in c}f(x)r(x).
\label{S:eq:restricted-cost}
\end{equation}
No renormalization of $G_c$ is needed: for every $x\in c$, all outputs with
$G(y|x)>0$ lie in $\Y_c$, and their probabilities still sum to one.

We next derive the mismatch decomposition within one island. Let $q^c$ be the
distribution that minimizes $C_c(G,f,r)$ over $r\in\Delta(c)$. For the cost
functions considered here, this minimizer has full support when $f(x)$ is
finite. To see this, suppose instead that a minimizing distribution $r$ gives
zero probability to at least one state. Because the island is connected,
there is a state $x_0$ outside the support of $r$ that shares at least one
possible output with a state in the support. Set
\begin{equation}
r_\epsilon=(1-\epsilon)r+\epsilon\delta_{x_0},
\qquad 0<\epsilon\ll1.
\end{equation}
Let
\begin{equation}
\alpha=\sum_{y:(G_cr)(y)=0}G(y|x_0).
\end{equation}
Since $x_0$ shares at least one output with the support of $r$, we have
$\alpha<1$. The new probability $\epsilon$ assigned to $x_0$ changes
$-S(r)$ by $-\epsilon\log(1/\epsilon)+O(\epsilon)$. New output
probabilities at outputs that had zero probability under $G_cr$ contribute
$\alpha\epsilon\log(1/\epsilon)+O(\epsilon)$ to $S(G_cr)$. All other
changes, including the $f$ term, are $O(\epsilon)$. Thus
\begin{equation}
C_c(G,f,r_\epsilon)-C_c(G,f,r)
=-(1-\alpha)\epsilon\log(1/\epsilon)+O(\epsilon)<0
\end{equation}
for sufficiently small $\epsilon$, contradicting the assumption that $r$
minimizes $C_c$. Therefore $q^c(x)>0$ for every $x\in c$.

For a full-support distribution, differentiating Eq.~\eqref{S:eq:restricted-cost}
with respect to $r(x)$ gives
\begin{equation}
\frac{\partial C_c}{\partial r(x)}
=\log r(x)-\sum_yG(y|x)\log(G_cr)(y)+f(x),
\label{S:eq:general-gradient}
\end{equation}
where the constant terms cancel because $\sum_yG(y|x)=1$. At the
minimizing distribution $q^c$, the normalization
condition $\sum_xq^c(x)=1$ requires these derivatives to have the same value
for all $x\in c$. Therefore
\begin{equation}
\log q^c(x)-\sum_yG(y|x)\log(G_cq^c)(y)+f(x)=\lambda_c,
\qquad x\in c,
\label{S:eq:general-stationarity}
\end{equation}
where $\lambda_c$ is independent of $x$.

Multiplying Eq.~\eqref{S:eq:general-stationarity} by $q^c(x)$ and summing
over $x$ gives
\begin{equation}
\lambda_c=C_c(G,f,q^c).
\label{S:eq:lambda-residual}
\end{equation}
For any distribution $r$ on $c$, Eq.~\eqref{S:eq:general-stationarity} also
gives
\begin{align}
&\KL{r}{q^c}-\KL{G_cr}{G_cq^c}\nonumber\\
&\quad=S(G_cr)-S(r)
+\sum_xr(x)\left[-\log q^c(x)
+\sum_yG(y|x)\log(G_cq^c)(y)\right]\nonumber\\
&\quad=S(G_cr)-S(r)+\sum_xf(x)r(x)-\lambda_c\nonumber\\
&\quad=C_c(G,f,r)-C_c(G,f,q^c).
\label{S:eq:restricted-mismatch-decomp}
\end{align}
Combining this identity with Eq.~\eqref{S:eq:cost-island-sum} gives
\begin{equation}
C(G,f,p)=\sum_cp(c)\left[R_c+
\MMC_c(p^c,q^c,G_c)\right],
\label{S:eq:cost-decomp}
\end{equation}
where
\begin{equation}
R_c=C_c(G,f,q^c)
\end{equation}
and
\begin{equation}
\MMC_c(r,q^c,G_c)
=\KL{r}{q^c}-\KL{G_cr}{G_cq^c}\ge0
\label{S:eq:mc}
\end{equation}
by the data-processing inequality.

The prior depends on the physical implementation through $f$. Conversely, any
desired full-support conditional prior can be obtained from a suitable choice
of $f$. For a chosen $q^c$, set
\begin{equation}
f_{q^c}(x)=-\log q^c(x)
+\sum_yG(y|x)\log(G_cq^c)(y),
\label{S:eq:f-from-q}
\end{equation}
up to an additive constant that is the same for every $x\in c$.
Equation~\eqref{S:eq:general-stationarity} is then satisfied by $q^c$.
Changing the prior while keeping $G$ fixed therefore corresponds to changing $f$.

Finally, no extra weights between island priors are needed in the mismatch
cost. Let $a_c>0$ with $\sum_ca_c=1$, and define
$q(x)=a_cq^c(x)$ for $x\in c$. Since the output identifies the island, the
same island weights $a_c$ appear in $Gq$. Directly separating the KL
divergences by island gives
\begin{align}
\KL{p}{q}
&=\sum_cp(c)\log\frac{p(c)}{a_c}
+\sum_cp(c)\KL{p^c}{q^c},\nonumber\\
\KL{Gp}{Gq}
&=\sum_cp(c)\log\frac{p(c)}{a_c}
+\sum_cp(c)\KL{G_cp^c}{G_cq^c}.
\label{S:eq:KL-islands}
\end{align}
The first terms are identical and cancel. Therefore
\begin{equation}
\MMC(p,\bm q,G)=\sum_cp(c)\MMC_c(p^c,q^c,G_c).
\label{S:eq:total-mmc}
\end{equation}

\section{Why a worst-case input can be a point mass}
\label{S:sec:pointmass}

Fix an island $c$ and a full-support candidate prior $q^c\in\Delta^\circ (c)$.
Define
\begin{equation}
d_c(x,q^c)=\KL{G(\cdot|x)}{G_cq^c},
\qquad
\varphi_c(x,q^c)=-\log q^c(x)-d_c(x,q^c).
\label{S:eq:phi}
\end{equation}
For any distribution $r\in\Delta(c)$, first note that
\begin{align}
&\sum_{x\in c}r(x)\KL{G(\cdot|x)}{G_cq^c}\nonumber\\
&\quad=\sum_{x,y}r(x)G(y|x)
\log\frac{G(y|x)}{(G_cq^c)(y)}\nonumber\\
&\quad=\sum_{x,y}r(x)G(y|x)
\log\frac{G(y|x)}{(G_cr)(y)}
+\sum_y(G_cr)(y)\log\frac{(G_cr)(y)}{(G_cq^c)(y)}\nonumber\\
&\quad=I_r(X;Y)+\KL{G_cr}{G_cq^c},
\label{S:eq:average-output-KL}
\end{align}
where $I_r(X;Y)$ is the input--output mutual information under $r$.

Using Eq.~\eqref{S:eq:average-output-KL},
\begin{align}
\sum_{x\in c}r(x)\varphi_c(x,q^c)
&=-\sum_xr(x)\log q^c(x)
-\sum_xr(x)\KL{G(\cdot|x)}{G_cq^c}\nonumber\\
&=H(r)+\KL{r}{q^c}
-I_r(X;Y)-\KL{G_cr}{G_cq^c}\nonumber\\
&=H_r(X|Y,J=c)+\MMC_c(r,q^c,G_c).
\label{S:eq:dual-identity}
\end{align}
The last equality uses
$H_r(X|Y,J=c)=H(r)-I_r(X;Y)$ within the fixed island $c$.

Since conditional entropy is nonnegative, Eq.~\eqref{S:eq:dual-identity}
gives, after bounding the average by the maximum,
\begin{equation}
\MMC_c(r,q^c,G_c)
\le\sum_xr(x)\varphi_c(x,q^c)
\le\max_{x\in c}\varphi_c(x,q^c).
\label{S:eq:pointmass-upper}
\end{equation}
For $r=\delta_x$ we have $H_r(X|Y,J=c)=0$, so
Eq.~\eqref{S:eq:dual-identity} gives
\begin{equation}
\MMC_c(\delta_x,q^c,G_c)=\varphi_c(x,q^c).
\label{S:eq:pointmass-equality}
\end{equation}
Choosing a state $x$ that maximizes $\varphi_c(x,q^c)$ reaches the upper
bound in Eq.~\eqref{S:eq:pointmass-upper}. Hence
\begin{equation}
F_c(q^c):=\max_{r\in\Delta(c)}\MMC_c(r,q^c,G_c)
=\max_{x\in c}\varphi_c(x,q^c).
\label{S:eq:pm}
\end{equation}
Thus, for every fixed full-support prior, at least one worst-case input is a
point mass. 

\section{The minimizing prior within one island and the global minimax value}
\label{S:sec:optimal}
Mathematically, a stochastic map $G$ is the same object as the transition
matrix of a discrete memoryless channel, and an island of $G$ corresponds to a
connected component of that channel. For a single island, the uniqueness and
full support of the maximizer of $H(X|Y)$, the fixed-point
equation~\eqref{S:eq:fp}, and the argument in
Eqs.~\eqref{S:eq:posterior-gap}--\eqref{S:eq:ratio-shared-output} below were
established by Nedi\'c \emph{et al.}~\cite{NedicEtAl2025} in the context of
commitment capacity. We give self-contained proofs because
Ref.~\cite{NedicEtAl2025} assumes that no input's output distribution equals,
or is a convex combination of, the output distributions of other inputs. We
don't make these assumptions: distinct input states of $G$ are physically distinct
and are kept even when their output distributions coincide. The new content of
this section is the identification of this maximizer with the minimax prior,
Eq.~\eqref{S:eq:local-dual}.
Define
\begin{equation}
h_c(r):=H_r(X|Y,J=c)
=H(r)+\sum_xr(x)H[G(\cdot|x)]-H(G_cr).
\label{S:eq:h}
\end{equation}
The second equality follows from
$H(X|Y)=H(X)+H(Y|X)-H(Y)$.

For a full-support distribution $r$, differentiating
Eq.~\eqref{S:eq:h} gives
\begin{align}
\frac{\partial h_c}{\partial r(x)}
&=-\log r(x)-1+H[G(\cdot|x)]
+\sum_yG(y|x)\bigl[\log(G_cr)(y)+1\bigr]\nonumber\\
&=-\log r(x)-\KL{G(\cdot|x)}{G_cr}\nonumber\\
&=\varphi_c(x,r).
\label{S:eq:grad-h}
\end{align}

We next show that $h_c$ has only one maximizing distribution. Let $r$ have
full support and let $v$ be any nonzero vector satisfying
$\sum_xv_x=0$. The second derivative of $h_c$ in the
direction $v$ is
\begin{equation}
v^\top\nabla^2h_c(r)v
=-\sum_{x\in c}\frac{v_x^2}{r(x)}
+\sum_{y\in\Y_c}
\frac{[\sum_{x\in c}G(y|x)v_x]^2}{(G_cr)(y)}.
\label{S:eq:hessian}
\end{equation}
For each output $y$, the inequality
\begin{equation}
\frac{[\sum_xG(y|x)v_x]^2}{(G_cr)(y)}
\le\sum_xG(y|x)\frac{v_x^2}{r(x)}
\label{S:eq:cs}
\end{equation}
follows by applying the Cauchy--Schwarz inequality to
$\sqrt{G(y|x)r(x)}$ and
$\sqrt{G(y|x)}v_x/\sqrt{r(x)}$. Summing
Eq.~\eqref{S:eq:cs} over $y$ and using $\sum_yG(y|x)=1$ gives
\begin{equation}
v^\top\nabla^2h_c(r)v\le0.
\end{equation}
Equality in Eq.~\eqref{S:eq:cs} for a given $y$ requires
$v_x/r(x)$ to have the same value for every $x$ with $G(y|x)>0$. If equality
held in the full sum, this condition would hold for every reachable output.
Because the states in one island are connected by shared outputs, it would
follow that $v_x/r(x)=\alpha$ for every $x\in c$. The condition
$\sum_xv_x=0$ then gives
$0=\alpha\sum_xr(x)=\alpha$, so $v=0$, contrary to the choice of $v$.
Therefore
\begin{equation}
v^\top\nabla^2h_c(r)v<0
\end{equation}
for every nonzero $v$ with $\sum_xv_x=0$. Thus $h_c$ is strictly concave
among full-support distributions.

It remains to rule out a maximizing distribution that gives zero probability
to one or more states. Suppose $r$ is such a maximizer. Since $c$ is one
island, there is a state $x_0$ outside the support of $r$ that shares at least
one possible output with a state in the support. Define
\begin{equation}
r_\epsilon=(1-\epsilon)r+\epsilon\delta_{x_0},
\qquad0<\epsilon\ll1,
\end{equation}
and let
\begin{equation}
A=\{y:(G_cr)(y)=0\},
\qquad
\alpha=\sum_{y\in A}G(y|x_0).
\end{equation}
At least one output of $x_0$ is already produced by the support of $r$, so
$\alpha<1$. We now expand the three terms in Eq.~\eqref{S:eq:h}. The new
input state $x_0$ gives
\begin{equation}
H(r_\epsilon)-H(r)
=\epsilon\log(1/\epsilon)+O(\epsilon).
\label{S:eq:input-boundary-expansion}
\end{equation}
The middle term of Eq.~\eqref{S:eq:h} is linear in $r$, so its change is
$O(\epsilon)$. For the output entropy, outputs in $A$ receive probability
$\epsilon G(y|x_0)$ and therefore contribute
\begin{equation}
H(G_cr_\epsilon)-H(G_cr)
=\alpha\epsilon\log(1/\epsilon)+O(\epsilon).
\label{S:eq:output-boundary-expansion}
\end{equation}
Outputs outside $A$ already have positive probability under $G_cr$, so their
entropy changes by only $O(\epsilon)$. Combining
Eqs.~\eqref{S:eq:input-boundary-expansion} and
\eqref{S:eq:output-boundary-expansion} gives
\begin{equation}
h_c(r_\epsilon)-h_c(r)
=(1-\alpha)\epsilon\log(1/\epsilon)+O(\epsilon)>0
\end{equation}
for sufficiently small $\epsilon$. This contradicts the assumption that $r$
maximizes $h_c$. Therefore every maximizer has full support. Since $h_c$ is
strictly concave among full-support distributions, the maximizer is unique.
Denote it by $q_c^*$.

At this maximum, the normalization condition $\sum_xr(x)=1$ requires the
partial derivatives in Eq.~\eqref{S:eq:grad-h} to have the same value for all
$x\in c$. Therefore, for some constant $\lambda_c$,
\begin{equation}
\varphi_c(x,q_c^*)=\lambda_c
\qquad\text{for every }x\in c.
\label{S:eq:equalize}
\end{equation}
Using the definition of $\varphi_c$ and exponentiating gives
\begin{equation}
q_c^*(x)=
\frac{\exp[-\KL{G(\cdot|x)}{G_cq_c^*}]}
{\sum_{z\in c}\exp[-\KL{G(\cdot|z)}{G_cq_c^*}]}.
\label{S:eq:fp}
\end{equation}
This is the self-consistency equation in the Letter.

We now show that the same distribution solves the minimax problem. For an
arbitrary full-support candidate prior $q^c$, use
Eq.~\eqref{S:eq:dual-identity} with $r=q_c^*$. Since
$F_c(q^c)$ is the largest point-mass cost, it is at least the average
point-mass cost under $q_c^*$, so
\begin{align}
F_c(q^c)
&\ge\sum_xq_c^*(x)\varphi_c(x,q^c)\nonumber\\
&=h_c(q_c^*)+\MMC_c(q_c^*,q^c,G_c)\nonumber\\
&\ge h_c(q_c^*).
\label{S:eq:minimax-lower}
\end{align}
For $q^c=q_c^*$, the mismatch term is zero. In addition,
Eq.~\eqref{S:eq:equalize} makes every point-mass cost equal to
$\lambda_c$, so its maximum and its average are both $\lambda_c$. Therefore
\begin{equation}
F_c(q_c^*)=h_c(q_c^*)=\lambda_c.
\label{S:eq:minimax-equality}
\end{equation}
Since $q_c^*$ maximizes $h_c$, Eqs.~\eqref{S:eq:minimax-lower} and
\eqref{S:eq:minimax-equality} give
\begin{equation}
F_c^*:=\min_{q^c}F_c(q^c)
=\max_{r\in\Delta(c)}H_r(X|Y,J=c).
\label{S:eq:local-dual}
\end{equation}

We also show that no other full-support prior has the same local minimax
value. Equality in Eq.~\eqref{S:eq:minimax-lower} would require
\begin{equation}
\MMC_c(q_c^*,q^c,G_c)=0.
\end{equation}
Define the joint distributions
\begin{equation}
P(x,y)=q_c^*(x)G(y|x),
\qquad
Q(x,y)=q^c(x)G(y|x).
\end{equation}
They have the same conditional distribution of $Y$ given $X$, so
\begin{equation}
\KL{P}{Q}=\KL{q_c^*}{q^c}.
\end{equation}
Writing the same joint KL divergence by conditioning on $Y$ gives
\begin{align}
\KL{q_c^*}{q^c}-\KL{G_cq_c^*}{G_cq^c}
&=\sum_y(G_cq_c^*)(y)
\KL{P(X|y)}{Q(X|y)}.
\label{S:eq:posterior-gap}
\end{align}
All terms on the right are nonnegative. Zero mismatch therefore requires
$P(X|y)=Q(X|y)$ for every reachable output $y$. For any $x$ with
$G(y|x)>0$, this equality gives
\begin{equation}
\frac{q_c^*(x)}{q^c(x)}
=\frac{(G_cq_c^*)(y)}{(G_cq^c)(y)}.
\label{S:eq:ratio-shared-output}
\end{equation}
Thus $q_c^*(x)/q^c(x)$ has the same value for all inputs that share an
output. The island is connected by chains of shared outputs, so this ratio is
constant over all $x\in c$. Since both distributions sum to one, the constant
must be one. Hence $q^c=q_c^*$.

We now combine the islands. For a collection $\bm q=\{q^c\}$,
Eq.~\eqref{S:eq:total-mmc} is a weighted average of the islandwise mismatch
costs. Therefore
\begin{equation}
F_G(\bm q):=\max_p\MMC(p,\bm q,G)
\le\max_cF_c(q^c).
\end{equation}
For an island that has the largest $F_c(q^c)$, choose a point-mass input on a
state that reaches the maximum in Eq.~\eqref{S:eq:pm}. This puts all input
probability on that island and makes the inequality an equality. Hence
\begin{equation}
F_G(\bm q)=\max_cF_c(q^c).
\label{S:eq:global-F}
\end{equation}
For every choice of priors,
$F_G(\bm q)\ge\max_cF_c^*$. Choosing $q^c=q_c^*$ in every island reaches
this lower bound, so
\begin{equation}
F_G^*=\max_cF_c^*.
\label{S:eq:global-minimax}
\end{equation}

Because the output determines the island label $J$,
\begin{equation}
H_p(X|Y)=\sum_cp(c)H_{p^c}(X|Y,J=c).
\label{S:eq:conditional-entropy-islands}
\end{equation}
For fixed island probabilities $p(c)$, the largest value is obtained by
maximizing the conditional entropy separately within each island. Using
Eq.~\eqref{S:eq:local-dual}, this gives
\begin{equation}
\max_{\{p^c\}}H_p(X|Y)=\sum_cp(c)F_c^*.
\end{equation}
The right side is a weighted average of the $F_c^*$. Its largest value is
$\max_cF_c^*$, obtained by putting all island probability on an island with
that value. Combining this with Eq.~\eqref{S:eq:global-minimax} gives
\begin{equation}
F_G^*=\max_{p\in\Delta(\X)}H_p(X|Y).
\label{S:eq:dual}
\end{equation}

For completeness, we characterize all globally minimizing collections of
priors. Let
\begin{equation}
M=\max_cF_c^*=F_G^*.
\end{equation}
A full-support collection $\bm q$ is a global minimizer exactly when
\begin{equation}
F_c(q^c)\le M
\qquad\text{for every }c.
\label{S:eq:global-optimizer-set}
\end{equation}
If this condition holds, Eq.~\eqref{S:eq:global-F} gives
$F_G(\bm q)\le M$, and $M$ is already the smallest possible global value, so
$F_G(\bm q)=M$. Conversely, if any island has $F_c(q^c)>M$, then
Eq.~\eqref{S:eq:global-F} gives $F_G(\bm q)>M$, so the collection is not a
global minimizer. If $F_c^*=M$, the local uniqueness proved above requires
$q^c=q_c^*$. If $F_c^*<M$, exact local minimization is not required; any
$q^c$ satisfying $F_c(q^c)\le M$ is allowed.

The quantity $\max_pH_p(X|Y)$ is called the maximum equivocation of a
communication channel. Winter, Nascimento, and Imai showed that, after
removing redundant input symbols, the commitment capacity of a discrete
memoryless channel equals this quantity~\cite{WinterNascimentoImai2003}.
Equation~\eqref{S:eq:dual} shows that the same conditional-entropy
maximization gives the minimax mismatch cost here. We keep physically
distinct input states even when they produce the same output distribution.

\section{Deterministic maps}
\label{S:sec:deterministic}

Let $G(y|x)=\mathbf1\{y=g(x)\}$ and let
$I_y=\{x:g(x)=y\}$. The nonempty sets $I_y$ are the islands. Within one
island, every input produces the same output, so
\begin{equation}
\MMC_y(r,q^y)=\KL{r}{q^y}.
\end{equation}
For any distribution $r$,
\begin{align}
\KL{r}{q^y}
&=-H(r)-\sum_xr(x)\log q^y(x)\nonumber\\
&\le -\sum_xr(x)\log q^y(x)\nonumber\\
&\le -\log\min_{x\in I_y}q^y(x).
\end{align}
A point mass on a state that has the smallest $q^y(x)$ reaches the last
bound. Therefore
\begin{equation}
F_y(q^y)=\max_r\KL{r}{q^y}
=-\log\min_{x\in I_y}q^y(x).
\end{equation}
Since $\min_{x\in I_y}q^y(x)\le 1/|I_y|$, with equality only for the
uniform distribution on $I_y$, the unique minimizer is uniform and
\begin{equation}
F_y^*=\log|I_y|.
\end{equation}
If
\begin{equation}
L_{\max}=\max_y|I_y|,
\end{equation}
then
\begin{equation}
F_G^*=\log L_{\max}.
\label{S:eq:det-result}
\end{equation}
The general global condition in Eq.~\eqref{S:eq:global-optimizer-set} becomes
\begin{equation}
-\log\min_{x\in I_y}q^y(x)\le\log L_{\max},
\end{equation}
or equivalently
\begin{equation}
\min_{x\in I_y}q^y(x)\ge L_{\max}^{-1}.
\end{equation}
For an island of size $L_{\max}$, normalization forces equality for every
state, so the prior must be uniform. Smaller islands can also satisfy the
condition with nonuniform priors.

\section{Computing the minimizing island prior}
\label{S:sec:algorithm}

For each island define
\begin{equation}
T_c(q^c)(x)=
\frac{\exp[-d_c(x,q^c)]}
{\sum_{z\in c}\exp[-d_c(z,q^c)]}.
\label{S:eq:T}
\end{equation}
Starting from any full-support $q_0^c$, define
\begin{equation}
q_{k+1}^c=T_c(q_k^c).
\end{equation}
Using Eq.~\eqref{S:eq:phi}, the same update can be written as
\begin{equation}
T_c(q^c)(x)=
\frac{q^c(x)e^{\varphi_c(x,q^c)}}
{\sum_{z\in c}q^c(z)e^{\varphi_c(z,q^c)}}.
\label{S:eq:T-phi}
\end{equation}
Every entry of $T_c(q^c)$ is positive, so the update remains full support.

We now show that one update cannot decrease $h_c$. Let
$T=T_c(q^c)$ and define
\begin{equation}
Z(q^c)=\sum_zq^c(z)e^{\varphi_c(z,q^c)}.
\end{equation}
Equation~\eqref{S:eq:T-phi} gives
\begin{equation}
\varphi_c(x,q^c)=\log\frac{T(x)}{q^c(x)}+\log Z(q^c).
\label{S:eq:phi-update}
\end{equation}
Using Eq.~\eqref{S:eq:dual-identity} first with $r=T$ and then with $r=q^c$
gives
\begin{align}
h_c(T)
&=\log Z(q^c)+\KL{G_cT}{G_cq^c},\nonumber\\
h_c(q^c)
&=\log Z(q^c)-\KL{q^c}{T}.
\label{S:eq:h-T-q}
\end{align}
Subtracting yields
\begin{equation}
h_c(T_c(q^c))-h_c(q^c)
=\KL{q^c}{T_c(q^c)}
+\KL{G_cT_c(q^c)}{G_cq^c}\ge0.
\label{S:eq:monotone}
\end{equation}
Equality requires $\KL{q^c}{T_c(q^c)}=0$, and therefore
$q^c=T_c(q^c)$. Thus the only points at which the update leaves $h_c$
unchanged are fixed points of $T_c$. A fixed point satisfies
Eq.~\eqref{S:eq:equalize}. The uniqueness proved in
Sec.~\ref{S:sec:optimal} therefore implies that the only fixed point is
$q_c^*$.

We now prove convergence directly. Fix a full-support $q^c$ and let
$T=T_c(q^c)$. Using Eq.~\eqref{S:eq:phi-update} and then
Eq.~\eqref{S:eq:dual-identity} with $r=q_c^*$,
\begin{align}
\KL{q_c^*}{q^c}-\KL{q_c^*}{T}
&=\sum_xq_c^*(x)\log\frac{T(x)}{q^c(x)}
=\sum_xq_c^*(x)\varphi_c(x,q^c)-\log Z(q^c)\nonumber\\
&=h_c(q_c^*)+\MMC_c(q_c^*,q^c,G_c)-\log Z(q^c).
\end{align}
The first line of Eq.~\eqref{S:eq:h-T-q} gives $\log Z(q^c)\le h_c(T)$.
Hence
\begin{equation}
\KL{q_c^*}{q^c}-\KL{q_c^*}{T_c(q^c)}
\ge h_c(q_c^*)-h_c(T_c(q^c))\ge0.
\label{S:eq:kl-descent}
\end{equation}
Apply Eq.~\eqref{S:eq:kl-descent} with $q^c=q_j^c$ and sum over
$j=0,\dots,k-1$. The left side telescopes, and by
Eq.~\eqref{S:eq:monotone} every term on the right is at least
$h_c(q_c^*)-h_c(q_k^c)$. Using $h_c(q_c^*)=F_c^*$,
\begin{equation}
0\le F_c^*-h_c(q_k^c)\le\frac{\KL{q_c^*}{q_0^c}}{k}.
\label{S:eq:rate}
\end{equation}
The right side is finite because $q_0^c$ has full support; for the uniform
start it is at most $\log|c|$. Thus $h_c(q_k^c)\to F_c^*$. Since $h_c$ is
continuous on the compact simplex $\Delta(c)$, every limit point $\bar q$ of
the sequence satisfies $h_c(\bar q)=F_c^*$, and the uniqueness of the
maximizer proved in Sec.~\ref{S:sec:optimal} gives $\bar q=q_c^*$. Because
every subsequence has a further subsequence converging to $q_c^*$,
\begin{equation}
q_k^c\longrightarrow q_c^*.
\label{S:eq:algorithm-convergence}
\end{equation}
This argument uses only Eqs.~\eqref{S:eq:dual-identity},
\eqref{S:eq:phi-update}, and \eqref{S:eq:monotone}, and it does not require
the channel to be nonredundant. The same iteration was analyzed in the
context of commitment capacity in Refs.~\cite{Hayashi2025,NedicEtAl2025}.

The local minimax value can be bounded at every step without waiting for full
convergence. For any full-support trial prior $q^c$,
Eq.~\eqref{S:eq:local-dual} gives
\begin{equation}
h_c(q^c)\le F_c^*,
\end{equation}
while the definition of $F_c^*$ gives
\begin{equation}
F_c^*\le F_c(q^c)=\max_{x\in c}\varphi_c(x,q^c).
\end{equation}
Therefore
\begin{equation}
h_c(q^c)\le F_c^*\le\max_{x\in c}\varphi_c(x,q^c).
\label{S:eq:local-bounds}
\end{equation}
Taking the largest lower and upper bounds over the islands gives
\begin{equation}
\max_ch_c(q^c)
\le F_G^*\le
\max_c\max_{x\in c}\varphi_c(x,q^c).
\label{S:eq:bounds}
\end{equation}
The difference between the two sides is therefore an upper bound on the
remaining error in the computed value of $F_G^*$.

\section{Additional consequences of the main result}
\label{S:sec:structural}

Several consequences follow directly from
$F_G^*=\max_pH_p(X|Y)$. Let $n=|\X|$ and define
\begin{equation}
C(G)=\max_pI_p(X;Y),
\end{equation}
the largest mutual information between input and output over all input
distributions. Using the uniform input distribution $u(x)=1/n$ gives
\begin{align}
F_G^*
&\ge H_u(X|Y)\nonumber\\
&=\log n-I_u(X;Y)\nonumber\\
&\ge\log n-C(G).
\end{align}
Also $H_p(X|Y)\le H_p(X)\le\log n$ for every $p$. Hence
\begin{equation}
\log n-C(G)\le F_G^*\le\log n.
\label{S:eq:general-bounds}
\end{equation}

If two inputs in the same island have identical output distributions,
$G(\cdot|x)=G(\cdot|x')$, then the two KL divergences in
Eq.~\eqref{S:eq:fp} are equal. Therefore
\begin{equation}
q_c^*(x)=q_c^*(x').
\end{equation}
More generally, suppose a relabeling of the input states together with a
relabeling of the output states leaves every transition probability
$G(y|x)$ unchanged. Applying that relabeling to $q_c^*$ produces another
maximizer of $h_c$. Since the maximizer is unique, the relabeled distribution
must equal $q_c^*$.

The minimax value is additive for independent product maps. Let
$G=G_1\otimes G_2$. For any joint input distribution,
\begin{align}
H(X_1,X_2|Y_1,Y_2)
&=H(X_1|Y_1,Y_2)
+H(X_2|X_1,Y_1,Y_2)\nonumber\\
&\le H(X_1|Y_1)+H(X_2|Y_2)\nonumber\\
&\le F_{G_1}^*+F_{G_2}^*,
\end{align}
where the first inequality holds because conditioning cannot increase
entropy. A product of input distributions that maximize
the two separate terms reaches the upper bound. Therefore
\begin{equation}
F_{G_1\otimes G_2}^*=F_{G_1}^*+F_{G_2}^*.
\end{equation}
If both maps have one island, their product also has one island. The product
of the two minimizing priors reaches the bound above, and the uniqueness
proved in Sec.~\ref{S:sec:optimal} then makes it the unique minimizing prior
for the product map.

\section{Driven three-state cycle}
\label{S:sec:ring}

Consider the stochastic map
\begin{equation}
G=\begin{pmatrix}
0.3&0.2&0.4\\
0.6&0.3&0.1\\
0.1&0.5&0.5
\end{pmatrix}.
\label{S:eq:ring-G}
\end{equation}
We use the column convention, so each column gives the output distribution for
one input state. Every entry is positive, so all three input states belong to
one island.

The stationary distribution is defined by $G\pi_{ss}=\pi_{ss}$. Solving these
equations together with $\sum_x\pi_{ss}(x)=1$ gives
\begin{equation}
\pi_{ss}=(0.306122\ldots,0.316327\ldots,0.377551\ldots).
\label{S:eq:ring-stationary}
\end{equation}
The minimax prior is instead the unique solution of
Eq.~\eqref{S:eq:fp}. Applying the update in
Sec.~\ref{S:sec:algorithm} gives
\begin{equation}
q^*=(0.300006\ldots,0.368165\ldots,0.331829\ldots).
\label{S:eq:ring-minimax}
\end{equation}
The two distributions are different:
\begin{equation}
\|q^*-\pi_{ss}\|_1\approx0.104.
\end{equation}

For this map,
\begin{equation}
F_G(q)=\max_x\left[-\log q(x)
-\KL{G(\cdot|x)}{Gq}\right].
\end{equation}
Evaluating this expression gives
\begin{equation}
F_G(q^*)\approx0.963,
\qquad
F_G(\pi_{ss})\approx1.112.
\end{equation}
Thus the stationary distribution does not minimize the worst-case mismatch
cost for one application of the map.

For reference, Eq.~\eqref{S:eq:ring-G} can be written as
\begin{equation}
G=\begin{pmatrix}
r_0&b_1&f_2\\
f_0&r_1&b_2\\
b_0&f_1&r_2
\end{pmatrix},
\end{equation}
with
\begin{align}
(f_0,b_0,r_0)&=(0.6,0.1,0.3),\nonumber\\
(f_1,b_1,r_1)&=(0.5,0.2,0.3),\nonumber\\
(f_2,b_2,r_2)&=(0.4,0.1,0.5).
\end{align}
The ratio of the product of the forward probabilities to the product of the
backward probabilities is
\begin{equation}
\frac{f_0f_1f_2}{b_0b_1b_2}=60,
\end{equation}
so the cycle affinity is $\mathcal A=\log60$. The stationary probability
current from state $0$ to state $1$ is
\begin{equation}
J_{ss}=f_0\pi_{ss}(0)-b_1\pi_{ss}(1)\approx0.120.
\end{equation}
The nonzero current shows that this example is not an equilibrium stationary
process.

Starting the algorithm from the uniform distribution, six updates give
\begin{equation}
0.9634104335\le F_G^*\le0.9634136415,
\end{equation}
and ten updates reduce the difference between the upper and lower bounds in
Eq.~\eqref{S:eq:bounds} below $5.5\times10^{-9}$ nats.

\section{A circuit versus the map it computes}
\label{S:sec:circuit}
A Boolean circuit is a model of computation that computes a boolean function by connecting simple logical gates, such as AND, OR, and NOT gates, so that the output of one gate can serve as the input to another. The same boolean function can therefore be treated either as a single map from the inputs to its final output, or as a sequence of local maps implemented by the individual gates.

\begin{figure}[t]
\centering
\includegraphics[width=0.72\textwidth]{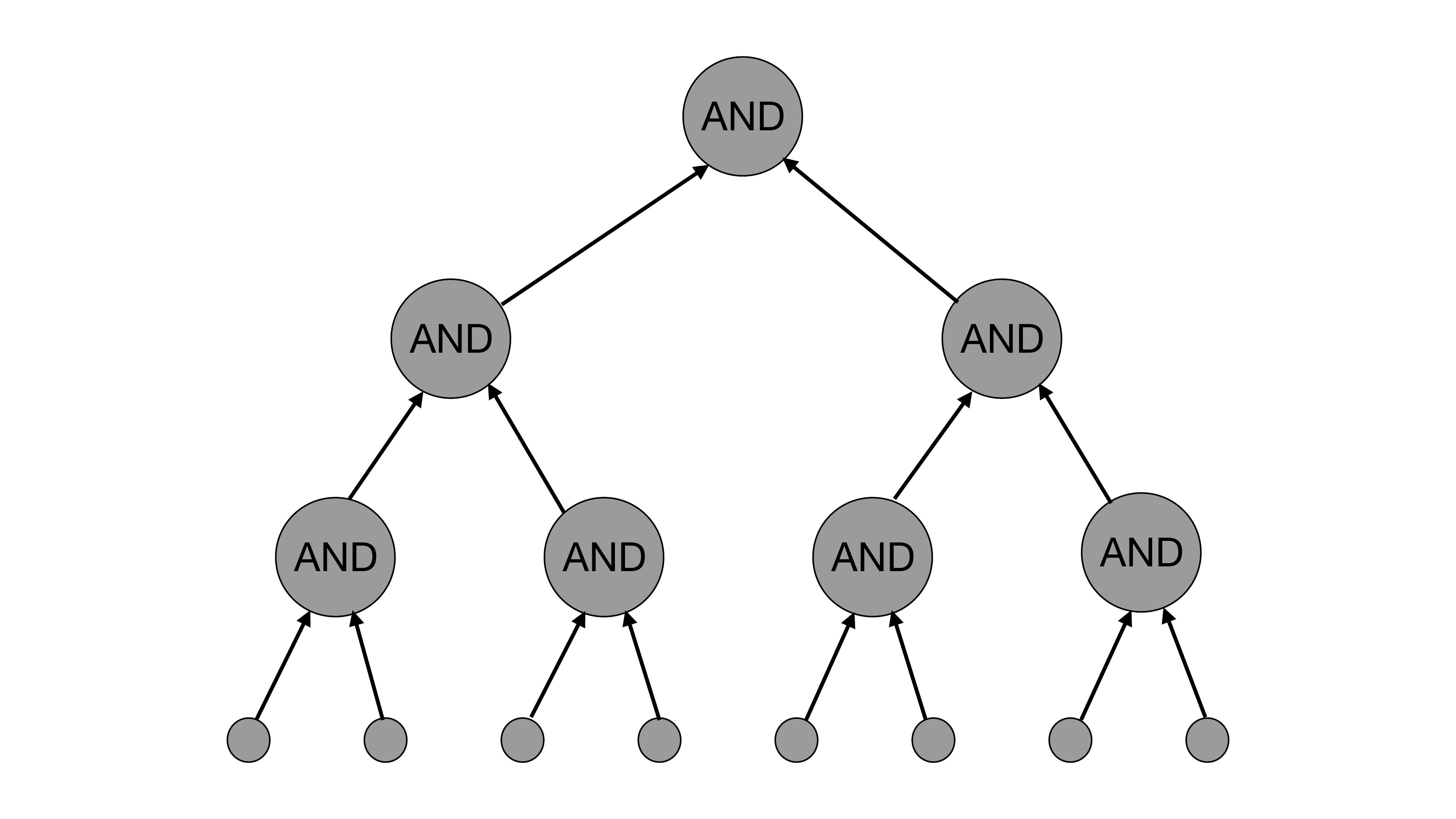}
\caption{A depth-$3$ binary tree of two-input AND gates. The eight bottom nodes represent the external input bits, and each higher node computes the AND of its two inputs. This figure first appeared in our previous work, Ref.~\cite{YadavYousefWolpert2025}.}
\label{S:fig:and-tree-illustration}
\end{figure}

Consider a complete binary tree of two-input AND gates of depth $d$. The
$2^d$ leaves are the external input bits, and each internal node applies an
AND operation to the two outputs immediately below it. There are therefore
$2^d-1$ gates in the tree. Figure~\ref{S:fig:and-tree-illustration} shows the
case $d=3$, with eight input bits and seven AND gates. The final output is the
AND of all $2^d$ input bits.

If the entire circuit is treated as one deterministic map from its $2^d$
input bits to the single final output bit, then the map has only two islands.
The output is $1$ only when every input bit is $1$, so the output-$1$ island
contains only the all-ones input. Every other one of the $2^{2^d}$ possible
input strings gives output $0$, so the output-$0$ island contains
$2^{2^d}-1$ states. Since the minimax mismatch cost of a deterministic map is
the logarithm of the size of its largest island, we obtain

\begin{equation}
F^*_{\rm map}(d)=\log(2^{2^d}-1)\approx2^d\log2.
\label{S:eq:tree-map}
\end{equation}

The corresponding minimax prior is uniform over the output-$0$ island, which
is the largest island.

If each AND gate is instead treated as
its own deterministic map. For one two-input AND gate, the output-$0$ island
is $\{00,01,10\}$ while the output-$1$ island is $\{11\}$. The largest island
therefore has three states, and the local minimax mismatch cost of one gate is
\begin{equation}
\log3.
\end{equation}

There are $2^d-1$ gates in the tree, so summing these local gate minimax costs
gives

\begin{equation}
F^*_{\rm gates}(d)=(2^d-1)\log3.
\label{S:eq:tree-gates}
\end{equation}

These two minimax costs 
refer to different physical decompositions of the same logical computation. For large $d$, the coarse-grained map cost approaches $2^d\log2$, whereas the
sum of the local gate costs approaches $2^d\log3$. Thus both scale linearly
with the number of gates, but with different coefficients.

Moreover, Eq.~\eqref{S:eq:tree-gates} includes only the sum of the local
mismatch costs of the individual gates. However, a gate's value in the AND tree
can be correlated with the rest of the circuit, and the total circuit
mismatch cost can contain additional mutual-information terms. Therefore the
local-gate sum in Eq.~\eqref{S:eq:tree-gates} is a lower bound on the full
circuit mismatch cost when those correlations are included, as discussed in our previous work~\cite{YadavYousefWolpert2025}.

\section{Instantaneous EP and mismatch cost}
\label{S:sec:continuous}

Let $P_t$ be a differentiable family of finite-state stochastic maps with
$P_0=I$, and define
\begin{equation}
A:=\left.\frac{dP_t}{dt}\right|_{t=0}.
\label{S:eq:A-def}
\end{equation}
Because every $P_t$ is stochastic and $P_0=I$, the columns of $A$ sum to zero
and $A_{yx}\ge0$ for $y\ne x$. We do not assume that the full dynamics are a
Markov semigroup, so in general we need not have $P_t=e^{tA}$. Let $f_t(x)$
be the expected entropy flow out of the system over $[0,t]$ when the initial
state is $x$, with $f_0(x)=0$. The accumulated EP is
\begin{equation}
W_d(t;p)=S(P_tp)-S(p)+p\cdot f_t.
\label{S:eq:accumulated-EP}
\end{equation}
For a full-support $p$, differentiability gives
\begin{equation}
P_tp=p+tAp+o(t).
\end{equation}
Therefore
\begin{align}
\left.\frac{d}{dt}S(P_tp)\right|_{t=0}
&=-\sum_x(Ap)(x)[\log p(x)+1]\nonumber\\
&=-\sum_x(Ap)(x)\log p(x),
\end{align}
because $\sum_x(Ap)(x)=0$. Differentiating
Eq.~\eqref{S:eq:accumulated-EP} gives
\begin{equation}
\sigma(p)=-(Ap)\cdot\ln p+p\cdot\dot f_0.
\label{S:eq:sigma}
\end{equation}

For full-support $p$, differentiating $\sigma(p)$ with respect to $p(x)$ gives
\begin{equation}
\frac{\partial\sigma}{\partial p(x)}
=-\sum_yA_{yx}\ln p(y)
-\frac{(Ap)(x)}{p(x)}+\dot f_0(x).
\label{S:eq:grad-sigma}
\end{equation}
Suppose the distribution $q^\sharp$ that minimizes $\sigma(p)$ has full
support. The normalization condition requires
\begin{equation}
-\sum_yA_{yx}\ln q^\sharp(y)
-\frac{(Aq^\sharp)(x)}{q^\sharp(x)}
+\dot f_0(x)=\mu
\label{S:eq:rate-prior}
\end{equation}
for every $x$. Multiplying this equation by $q^\sharp(x)$ and summing over
$x$ gives
\begin{equation}
\mu=\sigma(q^\sharp),
\end{equation}
because $\sum_x(Aq^\sharp)(x)=0$.

For arbitrary full-support $p$ and $q$, let
$a_t=P_tp$ and $b_t=P_tq$. Differentiating
$\KL{a_t}{b_t}=\sum_xa_t(x)\log[a_t(x)/b_t(x)]$ at $t=0$ gives
\begin{align}
\left.\frac{d}{dt}\KL{P_tp}{P_tq}\right|_{0}
&=\sum_x(Ap)(x)\log\frac{p(x)}{q(x)}
+\sum_x(Ap)(x)
-\sum_xp(x)\frac{(Aq)(x)}{q(x)}\nonumber\\
&=\sum_x(Ap)(x)\log\frac{p(x)}{q(x)}
-\sum_xp(x)\frac{(Aq)(x)}{q(x)}.
\label{S:eq:KL-rate}
\end{align}
The second line again uses $\sum_x(Ap)(x)=0$. We therefore define
\begin{equation}
\dot{\MMC}_A(p,q)
=-\sum_x(Ap)(x)\log\frac{p(x)}{q(x)}
+\sum_xp(x)\frac{(Aq)(x)}{q(x)}.
\label{S:eq:rate-explicit}
\end{equation}
Equivalently,
\begin{equation}
\dot{\MMC}_A(p,q)
=-\left.\frac{d}{dt}\KL{P_tp}{P_tq}\right|_{t=0}.
\end{equation}
This quantity is nonnegative because KL divergence cannot increase under a
stochastic map and $P_0=I$.

We now set $q=q^\sharp$. Multiply Eq.~\eqref{S:eq:rate-prior} by $p(x)$ and
sum over $x$:
\begin{equation}
-\sum_y(Ap)(y)\log q^\sharp(y)
-\sum_xp(x)\frac{(Aq^\sharp)(x)}{q^\sharp(x)}
+p\cdot\dot f_0
=\sigma(q^\sharp).
\label{S:eq:rate-prior-averaged}
\end{equation}
Subtracting this equation from Eq.~\eqref{S:eq:sigma} gives
\begin{align}
\sigma(p)-\sigma(q^\sharp)
&=-\sum_x(Ap)(x)\log\frac{p(x)}{q^\sharp(x)}
+\sum_xp(x)\frac{(Aq^\sharp)(x)}{q^\sharp(x)}\nonumber\\
&=\dot{\MMC}_A(p,q^\sharp)\nonumber\\
&=-\left.\frac{d}{dt}\KL{P_tp}{P_tq^\sharp}\right|_0.
\label{S:eq:rate-mmc}
\end{align}
This is the identity used in the Letter. Notice that only the derivative
$A=\dot P_0$ enters the instantaneous result; no Markov-semigroup assumption
is needed.

\section{The instantaneous minimax rate and the short-time finite-time limit}
\label{S:sec:shorttime}

We first prove the divergence of the instantaneous worst-case rate. Call the
initial dynamics nontrivial if at least one state has a positive escape rate.
Choose such a state $a$ and define
\begin{equation}
\lambda_a=\sum_{y\ne a}A_{ya}=-A_{aa}>0.
\label{S:eq:lambda-a}
\end{equation}
Let $n=|\X|$ and, for $0<\epsilon<1/n$, define the full-support distribution
\begin{equation}
p^{(\epsilon)}(a)=1-(n-1)\epsilon,
\qquad
p^{(\epsilon)}(x)=\epsilon\quad(x\ne a).
\label{S:eq:p-epsilon}
\end{equation}
For $x\ne a$,
\begin{equation}
(Ap^{(\epsilon)})(x)=A_{xa}+O(\epsilon),
\end{equation}
so
\begin{equation}
\sum_{x\ne a}(Ap^{(\epsilon)})(x)
=\lambda_a+O(\epsilon).
\label{S:eq:outflow-epsilon}
\end{equation}
For $x=a$,
$(Ap^{(\epsilon)})(a)=-\lambda_a+O(\epsilon)$.

Fix any full-support $q$. Using Eq.~\eqref{S:eq:rate-explicit}, separate the
terms that contain $\log p^{(\epsilon)}(x)$:
\begin{align}
\dot{\MMC}_A(p^{(\epsilon)},q)
&=-\sum_x(Ap^{(\epsilon)})(x)\log p^{(\epsilon)}(x)\nonumber\\
&\quad+\sum_x(Ap^{(\epsilon)})(x)\log q(x)
+\sum_xp^{(\epsilon)}(x)\frac{(Aq)(x)}{q(x)}.
\label{S:eq:rate-epsilon-split}
\end{align}
The second and third sums remain bounded as $\epsilon\to0^+$ because $q$ is
fixed and has full support. In the first sum, the $x=a$ term is $O(\epsilon)$
because $\log[1-(n-1)\epsilon]=O(\epsilon)$. For $x\ne a$,
$\log p^{(\epsilon)}(x)=\log\epsilon$. Using
Eq.~\eqref{S:eq:outflow-epsilon},
\begin{align}
-\sum_{x\ne a}(Ap^{(\epsilon)})(x)\log\epsilon
&=[\lambda_a+O(\epsilon)]\log(1/\epsilon)\nonumber\\
&=\lambda_a\log(1/\epsilon)+O(1).
\end{align}
Therefore
\begin{equation}
\dot{\MMC}_A(p^{(\epsilon)},q)
=\lambda_a\log(1/\epsilon)+O(1)
\qquad(\epsilon\to0^+).
\label{S:eq:rate-divergence-asymptotic}
\end{equation}
Since $\lambda_a>0$, the right side can be made arbitrarily large. Thus, for
every fixed full-support $q$,
\begin{equation}
\sup_{p\in\Delta^\circ (\X)}\dot{\MMC}_A(p,q)=+\infty,
\end{equation}
and consequently
\begin{equation}
\inf_{q\in\Delta^\circ (\X)}
\sup_{p\in\Delta^\circ (\X)}
\dot{\MMC}_A(p,q)=+\infty.
\label{S:eq:rate-infinite}
\end{equation}
The divergence comes from the supremum over distributions approaching a point
mass. For every fixed full-support pair $p,q$, all terms in
Eq.~\eqref{S:eq:rate-explicit} are finite.

\begin{figure}[t]
\centering
\includegraphics[width=0.55\textwidth]{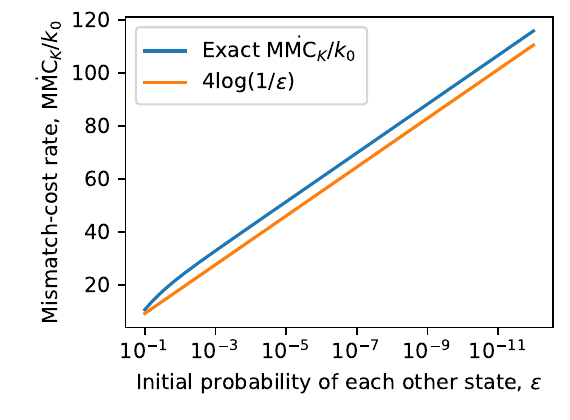}
\caption{Instantaneous mismatch-cost rate for the three-state nonequilibrium
steady state in Eq.~\eqref{S:eq:ness-K}, with $q=\pi_{\rm ss}$ and
$p^{(\epsilon)}$ given by Eq.~\eqref{S:eq:ness-peps}. The horizontal axis is
reversed so that moving to the right corresponds to $\epsilon\to0$. The exact
rate has the predicted $4\log(1/\epsilon)$ growth and therefore has no finite
upper bound.}
\label{S:fig:rate-divergence}
\end{figure}

As a concrete CTMC example, take $P_t=e^{tK}$ with
\begin{equation}
K=k_0
\begin{pmatrix}
-4 & 1/2 & 2/5\\
3 & -2 & 3/5\\
1 & 3/2 & -1
\end{pmatrix}.
\label{S:eq:ness-K}
\end{equation}
Its stationary distribution is
\begin{equation}
\pi_{\rm ss}=
\left(\frac{11}{112},\frac{9}{28},\frac{65}{112}\right).
\end{equation}
For example, the stationary current from state $0$ to state $1$ is
\begin{equation}
K_{10}\pi_{\rm ss}(0)-K_{01}\pi_{\rm ss}(1)
=\frac{15}{112}k_0,
\end{equation}
and the same nonzero current goes around the cycle. Thus the stationary state
is a nonequilibrium steady state. Set
\begin{equation}
q=\pi_{\rm ss},
\qquad
p^{(\epsilon)}=(1-2\epsilon,\epsilon,\epsilon).
\label{S:eq:ness-peps}
\end{equation}
Because $Kq=0$, Eq.~\eqref{S:eq:rate-explicit} with $A=K$ reduces to
\begin{equation}
\dot{\MMC}_K(p^{(\epsilon)},q)
=-\sum_x(Kp^{(\epsilon)})(x)
\log\frac{p^{(\epsilon)}(x)}{q(x)}.
\end{equation}
A direct multiplication gives
\begin{equation}
\frac{Kp^{(\epsilon)}}{k_0}
=
\left(
-4+\frac{89}{10}\epsilon,\,
3-\frac{37}{5}\epsilon,\,
1-\frac{3}{2}\epsilon
\right).
\end{equation}
The last two components approach $3k_0$ and $k_0$, respectively, while both
corresponding initial probabilities are $\epsilon$. Therefore
\begin{equation}
\frac{\dot{\MMC}_K(p^{(\epsilon)},q)}{k_0}
=
4\log(1/\epsilon)+O(1),
\qquad \epsilon\to0^+.
\label{S:eq:ness-divergence}
\end{equation}
Figure~\ref{S:fig:rate-divergence} compares the exact rate with
$4\log(1/\epsilon)$. The exact rate has the same logarithmic growth and
continues to increase as $\epsilon$ decreases, making the divergence visible.

To understand how this singular instantaneous behavior emerges from finite
times, we now study the minimax mismatch cost as the process duration approaches
zero. For this short-time analysis we specialize to a time-homogeneous CTMC
with $P_t=e^{tK}$. Assume $K$ is irreducible, meaning that every state can be
reached from every other state through a sequence of transitions with positive
rates. For a finite
irreducible continuous-time process, $P_t(y|x)>0$ for every $x,y$ and every
$t>0$. Therefore $P_t$ has one stochastic island. Equation~\eqref{S:eq:dual}
gives
\begin{equation}
F_t^*=\max_pH_p(X_0|X_t).
\label{S:eq:Ft-dual}
\end{equation}
Define
\begin{equation}
\lambda_x=-K_{xx}=\sum_{y\ne x}K_{yx},
\qquad
\lambda_{\max}=\max_x\lambda_x,
\qquad
\bar\lambda(p)=\sum_xp(x)\lambda_x.
\label{S:eq:escape-rates}
\end{equation}

We first prove an upper bound. Use $X_t$ itself as an estimate of the initial
state $X_0$. Its error probability is
\begin{align}
e_t(p)
&=\Pr[X_t\ne X_0]\nonumber\\
&=\sum_xp(x)[1-P_t(x|x)].
\end{align}
Since
\begin{equation}
P_t=I+tK+O(t^2),
\end{equation}
we have
\begin{equation}
1-P_t(x|x)=\lambda_xt+O(t^2).
\end{equation}
There are finitely many states, so the $O(t^2)$ bound can be chosen uniformly
in $x$ and therefore uniformly in $p$. Hence
\begin{equation}
e_t(p)=t\bar\lambda(p)+O(t^2).
\label{S:eq:error-probability}
\end{equation}
By Fano's inequality,
\begin{equation}
H(X_0|X_t)\le h_2(e)+e\log(n-1),
\label{S:eq:fano}
\end{equation}
where
\begin{equation}
h_2(e)=-e\log e-(1-e)\log(1-e)
\end{equation}
is the binary entropy. Applying Eq.~\eqref{S:eq:fano} to
Eq.~\eqref{S:eq:error-probability}, using
$\bar\lambda(p)\le\lambda_{\max}$ and
$h_2(e)=e\log(1/e)+O(e)$ as $e\to0$, gives
\begin{equation}
\limsup_{t\to0^+}
\frac{F_t^*}{t\log(\tau_0/t)}
\le\lambda_{\max}.
\label{S:eq:shorttime-upper}
\end{equation}
Here $\tau_0>0$ is any fixed reference time. Replacing $\log(1/t)$ by
$\log(\tau_0/t)$ changes only a term proportional to $t$, which does not
affect the limit.

For the lower bound, fix any full-support distribution $p$. The conditional
entropy can be written as
\begin{equation}
H_p(X_0|X_t)
=H(p)+\sum_xp(x)H(P_t(\cdot|x))-H(P_tp).
\label{S:eq:conditional-entropy-identity}
\end{equation}
For a fixed starting state $x$,
\begin{align}
P_t(x|x)&=1-\lambda_xt+O(t^2),\nonumber\\
P_t(y|x)&=tK_{yx}+O(t^2),\qquad y\ne x.
\label{S:eq:shorttime-transition}
\end{align}
For every $y\ne x$ with $K_{yx}>0$,
\begin{equation}
-[tK_{yx}+O(t^2)]\log[tK_{yx}+O(t^2)]
=tK_{yx}\log(1/t)+O(t).
\end{equation}
If $K_{yx}=0$, the corresponding probability is $O(t^2)$ and its entropy
contribution is $O(t^2\log(1/t))$, which is smaller than $O(t)$. The diagonal
term contributes only $O(t)$. Summing over $y$ therefore gives
\begin{equation}
H(P_t(\cdot|x))
=\lambda_xt\log(\tau_0/t)+O(t).
\label{S:eq:column-entropy}
\end{equation}
Because $p$ has full support,
$P_tp=p+tKp+O(t^2)$ stays in a region where all probabilities are positive
for sufficiently small $t$. The entropy is differentiable there, so
\begin{equation}
H(P_tp)=H(p)+O(t).
\label{S:eq:output-entropy-smooth}
\end{equation}
Substituting Eqs.~\eqref{S:eq:column-entropy} and
\eqref{S:eq:output-entropy-smooth} into
Eq.~\eqref{S:eq:conditional-entropy-identity} gives
\begin{equation}
H_p(X_0|X_t)
=\bar\lambda(p)t\log(\tau_0/t)+O_p(t),
\label{S:eq:fixed-p-shorttime}
\end{equation}
where the constant in $O_p(t)$ may depend on $p$.

For any $\eta>0$, choose a full-support distribution $p_\eta$ whose average
escape rate satisfies
\begin{equation}
\bar\lambda(p_\eta)\ge\lambda_{\max}-\eta.
\end{equation}
Such a distribution can be obtained by placing probability arbitrarily close
to one on a state with escape rate $\lambda_{\max}$ while keeping a positive
probability on every other state. Since
$F_t^*\ge H_{p_\eta}(X_0|X_t)$,
Eq.~\eqref{S:eq:fixed-p-shorttime} gives
\begin{equation}
\liminf_{t\to0^+}
\frac{F_t^*}{t\log(\tau_0/t)}
\ge\lambda_{\max}-\eta.
\end{equation}
Since this holds for every $\eta>0$,
\begin{equation}
\liminf_{t\to0^+}
\frac{F_t^*}{t\log(\tau_0/t)}
\ge\lambda_{\max}.
\label{S:eq:shorttime-lower}
\end{equation}
Combining Eqs.~\eqref{S:eq:shorttime-upper} and
\eqref{S:eq:shorttime-lower} proves
\begin{equation}
\lim_{t\to0^+}
\frac{F_t^*}{t\log(\tau_0/t)}
=\lambda_{\max}.
\label{S:eq:shorttime}
\end{equation}
Since $t\log(\tau_0/t)\to0$, this also shows $F_t^*\to0$, while
$F_t^*/t$ grows as $\lambda_{\max}\log(\tau_0/t)$.

We finish by proving the short-time behavior of the minimax prior. Let
$q_t^*$ be the minimax prior of $P_t$. Since $P_t$ has one island,
Eq.~\eqref{S:eq:Ft-dual} and the uniqueness result in
Sec.~\ref{S:sec:optimal} imply
\begin{equation}
F_t^*=H_{q_t^*}(X_0|X_t).
\end{equation}
Define
\begin{equation}
\bar\lambda(q_t^*)=\sum_xq_t^*(x)\lambda_x.
\end{equation}
We claim that
\begin{equation}
\bar\lambda(q_t^*)\longrightarrow\lambda_{\max}.
\label{S:eq:prior-average-rate}
\end{equation}
To prove this, suppose instead that there were a number $\delta>0$ and a
sequence $t_k\to0^+$ for which
$\bar\lambda(q_{t_k}^*)\le\lambda_{\max}-\delta$. The expansion
Eq.~\eqref{S:eq:error-probability} is uniform in the input distribution, so
along this sequence
\begin{equation}
e_{t_k}(q_{t_k}^*)
\le(\lambda_{\max}-\delta)t_k+O(t_k^2).
\end{equation}
Applying Eq.~\eqref{S:eq:fano} then gives
\begin{equation}
\limsup_{k\to\infty}
\frac{F_{t_k}^*}{t_k\log(\tau_0/t_k)}
\le\lambda_{\max}-\delta,
\end{equation}
which contradicts Eq.~\eqref{S:eq:shorttime}. Therefore
Eq.~\eqref{S:eq:prior-average-rate} holds.

Let
\begin{equation}
S_{\max}=\{x:\lambda_x=\lambda_{\max}\}.
\end{equation}
If every state is in $S_{\max}$, then
$q_t^*(S_{\max})=1$ for every $t$ and there is nothing more to prove.
Otherwise define
\begin{equation}
\gamma=\lambda_{\max}
-\max_{x\notin S_{\max}}\lambda_x>0.
\end{equation}
Then
\begin{align}
\lambda_{\max}-\bar\lambda(q_t^*)
&=\sum_xq_t^*(x)(\lambda_{\max}-\lambda_x)\nonumber\\
&\ge\gamma\sum_{x\notin S_{\max}}q_t^*(x)\nonumber\\
&=\gamma[1-q_t^*(S_{\max})].
\label{S:eq:prior-concentration}
\end{align}
The left side tends to zero by
Eq.~\eqref{S:eq:prior-average-rate}. Since $\gamma>0$, it follows that
\begin{equation}
q_t^*(S_{\max})\longrightarrow1.
\label{S:eq:prior-limit}
\end{equation}
If $S_{\max}$ contains only one state, this means that $q_t^*$ converges to
the point-mass distribution on that state.

\end{document}